\documentclass{article}

\usepackage[a4paper,margin=25mm]{geometry}
\usepackage{microtype}
\usepackage{graphicx}
\usepackage{amsmath,amssymb,amsfonts}
\usepackage{bm}
\usepackage{physics}
\usepackage{siunitx}
\usepackage{authblk}
\usepackage[numbers]{natbib}
\usepackage[colorlinks=true,linkcolor=blue,citecolor=blue,urlcolor=blue]{hyperref}
\usepackage{caption}
\usepackage{subcaption}
\usepackage{booktabs}
\usepackage{placeins}
\usepackage{xcolor}
\usepackage{indentfirst}
\usepackage{float}
\usepackage{overpic}

\title{Generation of high-OAM ultraviolet twisted light for RF-photoinjector applications\thanks{This document is the results of the research project funded by the Russian Science Foundation (Project No. 23-62-10026)~\cite{RSF}}}
\author{A.S.~Dyatlov\textsuperscript{1,2,\thanks{Email: \href{mailto:aleksandr.dyatlov@metalab.ifmo.ru}{aleksandr.dyatlov@metalab.ifmo.ru}}}, 
D.M.~Dolgintsev\textsuperscript{1}, 
V.V.~Gerasimov\textsuperscript{3,4}, 
V.V.~Kobets\textsuperscript{2}, 
V.P.~Nazmov\textsuperscript{3,5},
M.A.~Nozdrin\textsuperscript{2},
A.N.~Sergeev\textsuperscript{1},
D.S.~Shokin\textsuperscript{2},
K.E.~Yunenko\textsuperscript{2},
and D.V.~Karlovets\textsuperscript{1,6}
}

\affil{\textit{\textsuperscript{1} School of Physics and Engineering, ITMO University, 9 Lomonosova St., Saint-Petersburg, 191002, Russia}}
\affil{\textit{\textsuperscript{2} Joint Institute for Nuclear Research, 6 Joliot-Curie St., Dubna, 141980, Moscow Region, Russia}}
\affil{\textit{\textsuperscript{3} Budker Institute of Nuclear Physics, Siberian Branch of RAS, 11, Acad. Lavrentieva Pr., Novosibirsk, 630090, Russia}}
\affil{\textit{\textsuperscript{4} Department of Physics, Novosibirsk State University, 1, Pirogova St., Novosibirsk, 630090, Russia}}
\affil{\textit{\textsuperscript{5} Institute of Solid State Chemistry and Mechanochemistry, Siberian Branch of RAS, 18, Kutateladze St., Novosibirsk, 630090, Russia}}
\affil{\textit{\textsuperscript{6} Petersburg Nuclear Physics Institute of NRC "Kurchatov Institute", 1, mkr. Orlova roshcha, Gatchina, Leningradskaya Oblast, 188300, Russia}}

\begin{document}

\date{}
\maketitle

\begin{abstract}
    The generation of relativistic vortex electron beams via photoemission requires ultraviolet (UV) laser beams with well-controlled orbital angular momentum (OAM) 
    compatible with radio-frequency (RF) photoinjector drive-laser systems. However, achieving high-OAM beam generation in the deep UV with sufficient efficiency, stability, 
    and mode control remains technically challenging. Here, we experimentally demonstrate the generation of high-OAM (up to $\ell = 64\hbar$) UV vortex beams at 266 nm 
    using three types of fabricated diffractive optical elements integrated into an operational photoinjector drive-laser system: a reflective fork grating, a 
    high–topological-charge spiral phase plate, and binary axicons. The spiral phase plate produces a high-purity Laguerre–Gaussian mode with a conversion efficiency of 80\%, 
    while fork gratings provide flexible access to lower-order OAM states and enable robust modal diagnostics. In contrast, binary axicons generate low-divergence quasi–Bessel 
    beams that can be interpreted as controlled superpositions of multiple OAM states with a finite OAM bandwidth. The generated beams are characterized using 
    cylindrical-lens mode conversion and radial intensity analysis, demonstrating controlled generation of both near-pure OAM eigenstates and broadband OAM distributions in the 
    UV regime. These results provide a comparative, application-driven framework for selecting UV-compatible OAM generation techniques and establish a practical route toward 
    structured photocathode illumination in high-brightness RF photoinjectors.
\end{abstract}

\section{Introduction}\label{sec:introduction}
    Optical beams of the so-called twisted light~\cite{Allen92, AndrewsBabiker12, Franke-Arnold08,Bliokh17, Babiker19, Floettmann20, Karlovets21, Ivanov22, Fu24}, carrying 
    orbital angular momentum (OAM), play an increasingly important role in modern photonics, enabling advanced control of light--matter interactions~\cite{Schimmoller24}, 
    high-dimensional quantum communication protocols~\cite{Brandt20, Fickler20, Cao20}, and structured-beam excitation in electron and plasma systems~\cite{Denoeud17}. 
    Generating well-defined OAM modes in the deep ultraviolet (UV) range is particularly relevant for photocathode-based electron sources and accelerator-driven applications, 
    where high spatial coherence, low divergence, and precise transverse-mode control are required.    

    However, producing stable high-charge OAM beams in the deep UV remains technically challenging. Spatial light modulators and digital micromirror devices, widely used at 
    visible wavelengths, suffer from strong absorption, laser-induced damage, and low diffraction efficiency in the UV range. Therefore, robust diffractive optical 
    elements are highly demanded for high-power applications, such as driving an RF photoinjector. While the generation of extremely high-order OAM beams is routinely 
    demonstrated at visible and infrared wavelengths, our work directly addresses the specific material limitations of the deep UV band, demonstrating a reliable, damage-resistant 
    approach. Binary spiral zone plates (SZPs) operate in this spectral range but are intrinsically limited to a conversion efficiency of about 40\%~\cite{Dyatlov26}. Fork gratings provide 
    efficient generation of OAM modes with flexible 
    control via diffraction order but restrict accessible values to discrete multiples and typically require tight focusing~\cite{Solomonov24}. Spiral phase plates (SPPs), 
    while conceptually ideal for generating arbitrary OAM values~\cite{Khonina20}, demand nanometer-scale surface accuracy over millimeter-sized apertures~\cite{Oemrawsingh04} 
    and must withstand significant UV fluences. Binary axicons, widely used to generate Bessel-like vortex beams~\cite{Osintseva22, Knyazev23, Bazdyrev25}, are robust and tunable
    but introduce characteristic ring fragmentation due to their discretized phase profiles. As a consequence, none of these approaches alone provides a 
    universal, high-fidelity, UV-compatible solution for high-order OAM generation. Instead, the existing methods involve trade-offs between efficiency, mode purity, achievable 
    OAM values, and robustness under UV irradiation, and no single approach is optimal under realistic photoinjector conditions.

    In this work, we demonstrate the generation of deep-UV
    twisted light using three fabricated diffractive optical elements (DOEs): reflective fork gratings, a 
    high-topological-charge SPP,
    and binary axicons, all integrated into an RF photoinjector drive-laser system~\cite{Gacheva14}. We generate Laguerre--Gaussian 
    (LG) modes with OAM values up to $\ell = 64 \hbar$ 
    at a wavelength of 266~nm and experimentally assess the performance of each approach in terms of mode purity, divergence, 
    and operational robustness. The modal structure is verified using cylindrical-lens mode conversion, and the measured intensity distributions are compared with numerical 
    simulations. In contrast, binary axicons generate low-divergence quasi--Bessel beams that can be interpreted as controlled superpositions of multiple OAM states, resulting 
    in a finite and tunable OAM spectral bandwidth. The generated beams exhibit high spatial fidelity, stable operation under UV irradiation, and low divergence suitable for 
    photocathode illumination. The fabrication techniques used here rely on robust, scalable processes compatible with fused silica, SU-8 resists, and metal-coated substrates, 
    enabling durable large-aperture UV optics. Therefore, this work
    provides a comparative, application-driven evaluation of UV-compatible DOEs for high-OAM beam generation under 
    realistic
    operating conditions of an RF photoinjector.

    To the best of our knowledge, this is the first experimental realization of deep-UV high-OAM beams based on multiple types of fabricated DOEs integrated into a single RF 
    photoinjector drive-laser system and validated through mode-conversion diagnostics. The presented results are directly relevant to accelerator facilities employing vortex 
    electron beams,
    offering an additional degree of freedom complementary to spin-polarized beams for probing spin-dependent phenomena~\cite{Floettmann20, Ivanov22, 
    SchattschneiderPRB12, Wu19, An25}.

    The demonstrated approach to deep-UV high-OAM beam generation is
    a practical and scalable solution for implementing controlled OAM states in high-brightness 
    photoinjectors under realistic UV operating conditions. This capability is directly relevant for accelerator-based light sources, including free-electron lasers, 
    synchrotron radiation facilities, and Compton $\gamma$-ray sources, where precise control of the transverse beam structure and phase space is required for advanced beam 
    manipulation and optimization.

\section{Design and fabrication of high-OAM diffractive optical elements}\label{sec:design}

\subsection{A reflective diffraction fork grating with a topological charge m~=~2}\label{sec:fork}

    A fork grating introduces a phase dislocation with a
    topological charge $m$, generating optical vortices in the diffracted orders. The phase profile employed in this work 
    follows Ref.~\cite{Allen92}:

    \begin{equation}\label{eq:fork_grat_func}
        \Phi(x,y) = 2\pi \frac{x}{x_0} + m\,\arctan{\frac{y}{x}},
    \end{equation}
    where $x_0$ denotes the grating period. The continuous phase distribution is converted into a binary mask by applying a cosine-based threshold function,     
    
    \begin{equation}
        B(x, y) = 
        \begin{cases} 
            1, & M(x,y) > T,\\ 
            0, & M(x,y) \le T. 
        \end{cases}
    \end{equation}
    with the modulation function defined as

    \begin{equation}
        M(x,y) = \frac{1}{2}\,\alpha[1 + \cos \Phi(x,y)].
    \end{equation}
    
    In this context, the topological charge of the grating $m$ is the number of $2\pi$ phase windings embedded into the grating period. When a Gaussian beam is diffracted, 
    each diffraction order acquires an optical vortex with the topological charge of the beam~\cite{Kotlyar22}
    
    \begin{equation}
    \ell = n \cdot m \hbar.
    \end{equation}
    where $n$ is the diffraction order. 
    The vortex charge $\ell$ defines the OAM carried by the beam, with each photon transferring an OAM of $\ell \hbar$.

\begin{figure}[!ht]
        \centering
        \begin{minipage}[t]{0.49\linewidth}
            \centering
            \includegraphics[width=\linewidth]{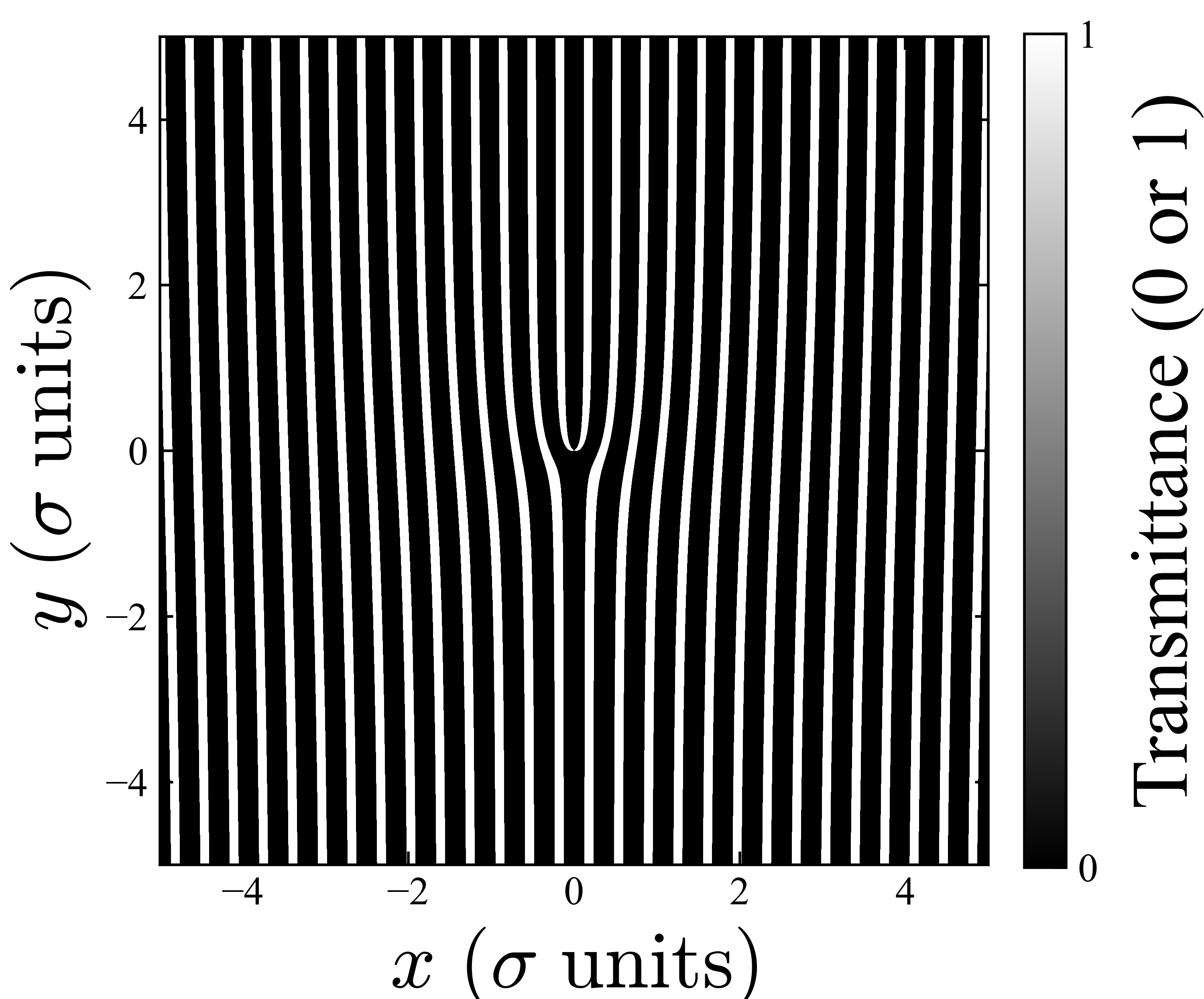}
            \caption{Binarized fork-grating mask defined by Eq.~\ref{eq:fork_grat_func}. Grayscale indicates binary transmittance (0 or 1) in normalized units.}            
            \label{fig:fork_grat_mask}
            \end{minipage}\hfill
        \begin{minipage}[t]{0.47\linewidth}
            \centering
            \includegraphics[width=0.85\linewidth]{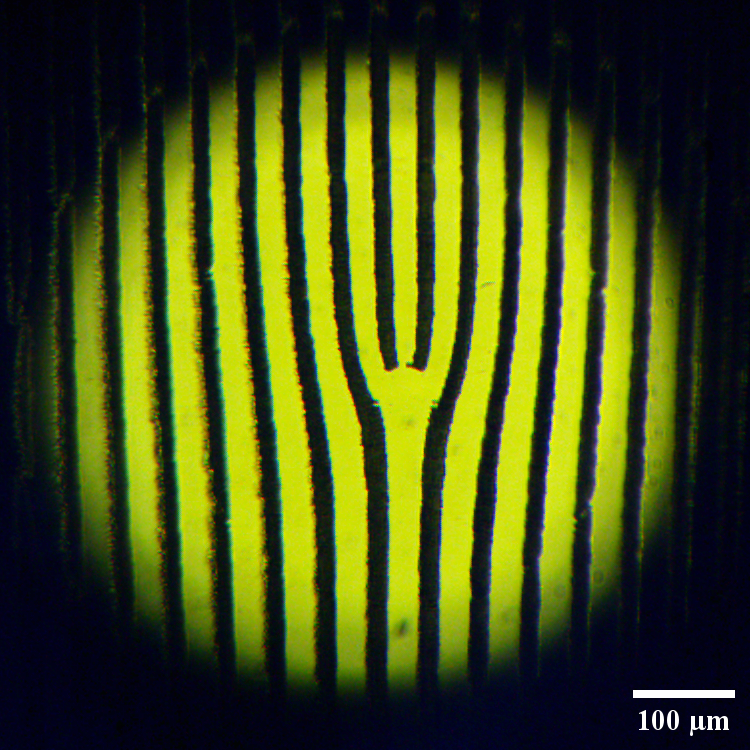}
            \caption{Microscope image of the laser-engraved reflective fork grating on aluminum, replicating the mask (a). Active area: $1 \times 1~\mathrm{mm}$; resolution: $30~\mu\mathrm{m}$.}
            \label{fig:fork_grat}
        \end{minipage}
    \end{figure}

    The resulting binary mask is shown in Figure~\ref{fig:fork_grat_mask}. The pattern was transferred onto an aluminum mirror via laser engraving, yielding a reflective 
    grating with a $1 \times 1~\mathrm{mm}$ active area and a spatial resolution of approximately $30~\mu\mathrm{m}$. A microscope image of the fabricated device is presented 
    in Figure~\ref{fig:fork_grat}.
    
\subsection{A spiral phase plate with a topological charge m~=~64}\label{sec:spp}

    A spiral phase plate (SPP) generates an azimuthally varying phase through a helical surface relief. The phase shift introduced by the plate is given by~\cite{Oemrawsingh04}

    \begin{equation}\label{eq:spp_func}
        \varphi_0(\theta) = \frac{2\pi}{\lambda} \left[ (n - n_0)\frac{h_s \theta}{2\pi} + n h_0 \right],
    \end{equation}
    where $n$ and $n_0$ are the refractive indices of the plate material and the surrounding medium, respectively; $h_0$ is the base thickness; $h_s$ is the total height 
    variation over a full $2\pi$ rotation; and $\theta$ is the azimuthal angle. The corresponding surface profile is

    \begin{equation}
        h(\theta) = h_s \frac{\theta}{2\pi} + h_0.
    \end{equation}

    To generate
    twisted light with an OAM of $\ell$, the phase must increase by $2\pi m$ over one full revolution:

    \begin{equation}
        \varphi_0(\theta) = \ell \theta + \frac{2\pi n h_0}{\lambda}.
    \end{equation}

\begin{figure}[!ht]
        \centering
        \begin{minipage}{0.48\linewidth}
            \centering
            \includegraphics[width=\linewidth]{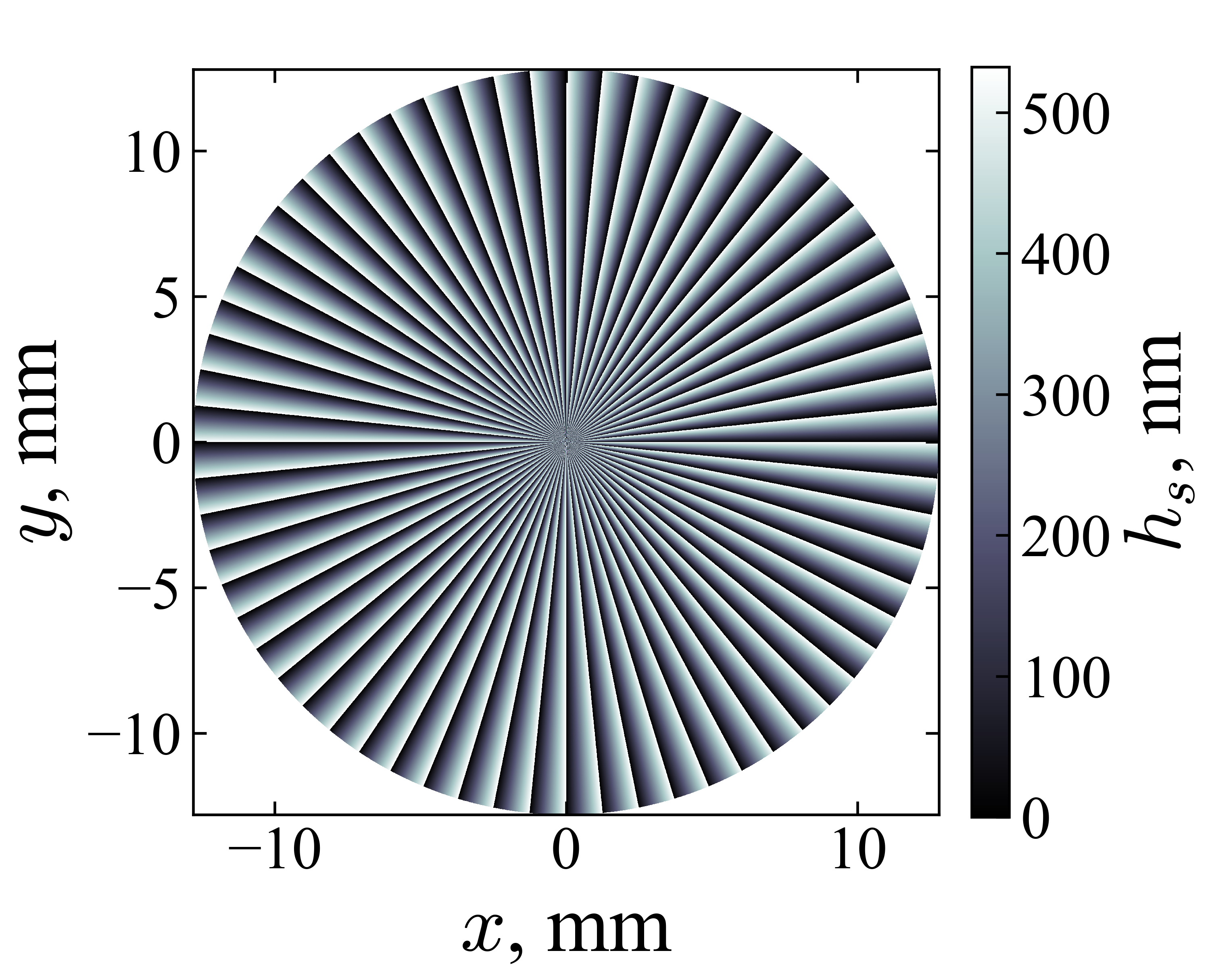}
            \caption{2D phase topology of the SPP ($m=64$) with 64 azimuthal steps and aperture $D = 25.6\ \mathrm{mm}$. 
            Calculated step height is $h_s \approx 532\ \mathrm{nm}$ for $\lambda = 266\ \mathrm{nm}$ and $n = 1.49$.}
            \label{fig:2D_map_spp}
        \end{minipage}\hfill
        \begin{minipage}{0.48\linewidth}
            \centering
            \includegraphics[width=0.9\linewidth]{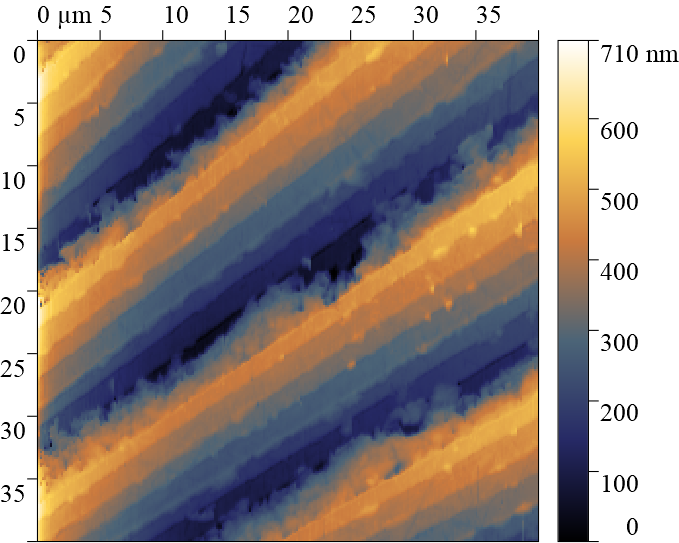}
            \caption{AFM image of the fabricated SPP. Minor surface dust and etching patterns do not degrade the beam quality. Measured step height $\sim 500\ \mathrm{nm}$ matches the design.}            
            \label{fig:spp_structure}
        \end{minipage}
    \end{figure}

    For UV operation at $\lambda = 266~\mathrm{nm}$ using fused silica ($n = 1.49$),
    a total height variation
    of approximately $h_s \approx 34~\mu\mathrm{m}$ is required. 
    A topological charge of $m = 64$ is achieved by dividing the spiral into 64 discrete azimuthal sectors, each introducing a $2\pi$ phase step. This corresponds 
    to a height increment of about $532~\mathrm{nm}$ per sector. The resulting two-dimensional phase topology is shown in Figure~\ref{fig:2D_map_spp}.

    The SPP was fabricated in fused silica using multi-level photolithography followed by high-precision plasma etching (Holo/Or, Israel). This is a common manufacturing process
    for diffractive optical elements, providing
    the required micron-scale surface relief and sub-micron accuracy of the phase profile, and ensuring stable 
    performance under high-intensity UV illumination~\cite{Goebel96, Wang97, Kostyuk22, Shi24}. 
    
    Figure~\ref{fig:spp_structure} shows the
    surface profile of the fabricated SPP
    measured using an AIST-NT SmartSPM~1000 atomic force microscope (AFM). The AFM image 
    reveals
    small dust particles accumulated inside the narrow gaps between adjacent sectors during handling, although this contamination did not affect the quality of the 
    generated twisted light. A weak striped pattern is also visible across the sectors, originating from layer-by-layer etching of 
    the fused-silica substrate. Despite
    these negligible stripes, the average step height remains close to 500~nm, which is in good agreement with the designed two-dimensional topological phase map.

\subsection{Binary axicons with the topological charge m~=~3 and 10}\label{sec:axicons}

    \begin{figure*}[!ht]
        \centering
        \includegraphics[width=0.7\linewidth]{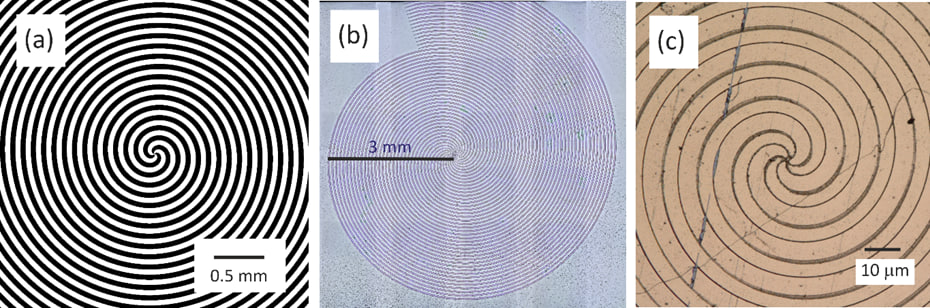}
        \caption{(a) Phase mask of an axicon with topological charge  $m$ = 3. 
        Micrograph of a binary planar axicon with an aperture diameter of 6 mm and $m$ = 3 in a SU-8 photoresist layer; (b) full view; (c) enlarged view obtained in 
        refractive mode, covered with a gold layer}
        \label{fig:axicons_mask}
    \end{figure*}

    Bessel beams were modeled using the Wave-ThruMasks MATLAB package~\cite{Kameshkov19}, which computes field propagation through transparent phase elements within the scalar 
    diffraction theory using the Rayleigh--Sommerfeld integral~\cite{Goodman05}. The binary phase distribution used to generate twisted light -- Bessel beams is

    \begin{equation}
        H(r, \varphi) = F(\ell \varphi - kr) = \frac{\pi}{2} \times \text{sign} \left[\sin(m \varphi - kr) \right],
    \end{equation}
    where $m$ is the topological charge, $\varphi$ the azimuthal angle, $k$ the radial wave number, and $r = \sqrt{x^2 + y^2}$. 
    An example mask for $m = 3$ is shown in Figure~\ref{fig:axicons_mask}(a).

    Binary axicon topologies in the form of Archimedean spirals with $m = 3$ and $m = 10$ and a period of $p = 100~\mu\mathrm{m}$ were generated in MathCad. The rendering 
    process was terminated once the spirals reached the working aperture of 6~mm.

    The encoded topology was first transferred onto a glass photomask in an iron-oxide layer. The axicons were then fabricated via photolithography using
    SU-8 photoresist 
    deposited on leucosapphire substrates, which provide high UV
    transmittance. The polymer crosslinked network of SU-8 also offers high resistance to ionizing 
    radiation~\cite{Nazmov04}. Prior to resist deposition, the substrates were cleaned sequentially in acetone and in a 5\% aqueous sulfuric-acid solution.
    To achieve the 
    required film thickness, SU-8 2002 resist was diluted with an SU-8 2000 thinner.

    The required binary layer thickness is

    \begin{equation}
        h = \frac{\lambda}{2 (n-1)},
    \end{equation}
    where $n \approx 1.66$ is the refractive index of SU-8. For $\lambda = 266~\mathrm{nm}$, this yields $h \approx 200~\mathrm{nm}$, providing the necessary $\pi$ phase shift.

    The photoresist was 
    spin-coated at 3000~rpm. The resulting film was soft-baked on a hotplate at $95^{\circ}\mathrm{C}$ for 5 minutes. The axicon 
    pattern was transferred onto the photoresist using a custom UV exposure system based on an H44TV1C0-LFVY LED lamp operating at a wavelength of 365~nm, with an exposure time of 23 minutes. After 
    a post-exposure bake at $95^{\circ}\mathrm{C}$ for 5 minutes and development for 1.5 minutes in PGMEA, the resist thickness was measured with a Linnik MII-4 interference 
    microscope with an accuracy of $\pm 15~\mathrm{nm}$. Figures~\ref{fig:axicons_mask}(b--c) show the resulting axicon relief.

\section{Numerical simulation of the generated beams}\label{sec:num_sim}

    The fork grating and the SPP
    were numerically simulated using a standard angular-spectrum propagation framework~\cite{Goodman05}. 
    A Gaussian input beam was assumed and multiplied by the corresponding phase profile of the diffractive optical element: $\Phi(x,y)$ for the fork grating 
    (Eq.~\ref{eq:fork_grat_func}) and $\varphi_0(\theta)$ for the SPP (Eq.~\ref{eq:spp_func}). Free-space propagation and lens transformations were implemented using Fourier-domain 
    transfer operators.

    To suppress boundary effects and residual on-axis intensity that become pronounced for high-OAM modes, a radial apodization was applied, introducing an outer truncation and 
    a central suppression region. The specific functional form was chosen to minimize numerical artifacts without affecting the intrinsic transverse mode structure.
   
\begin{figure}[!ht]
      \centering
      \includegraphics[width=0.7\linewidth]{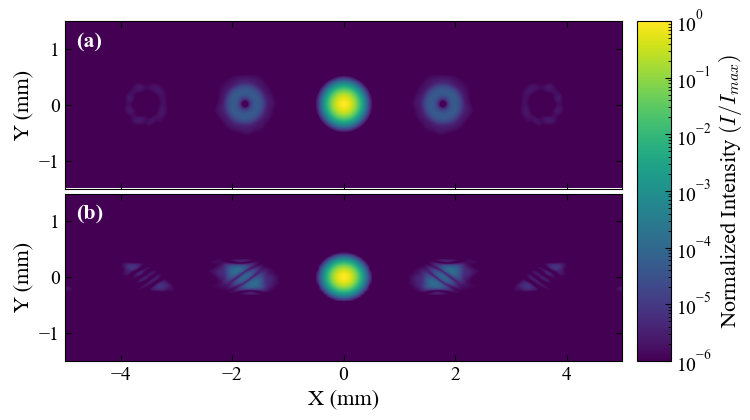}
      \caption{Simulation of
      twisted light generation using the fork grating.  
      \textbf{(a)} Transformation of the Gaussian input into the Laguerre--Gaussian modes carrying the OAMs $\ell = 0 \hbar,\ \pm 2 \hbar,\ \pm 4 \hbar$.  
      \textbf{(b)} Corresponding cylindrical-lens mode conversion into the Hermite--Gaussian modes of the orders $N = 1,\ 3,\ 5$.}
      \label{fig:fork_sim}
    \end{figure}

\begin{figure*}
        \centering
        \includegraphics[width=0.8\linewidth]{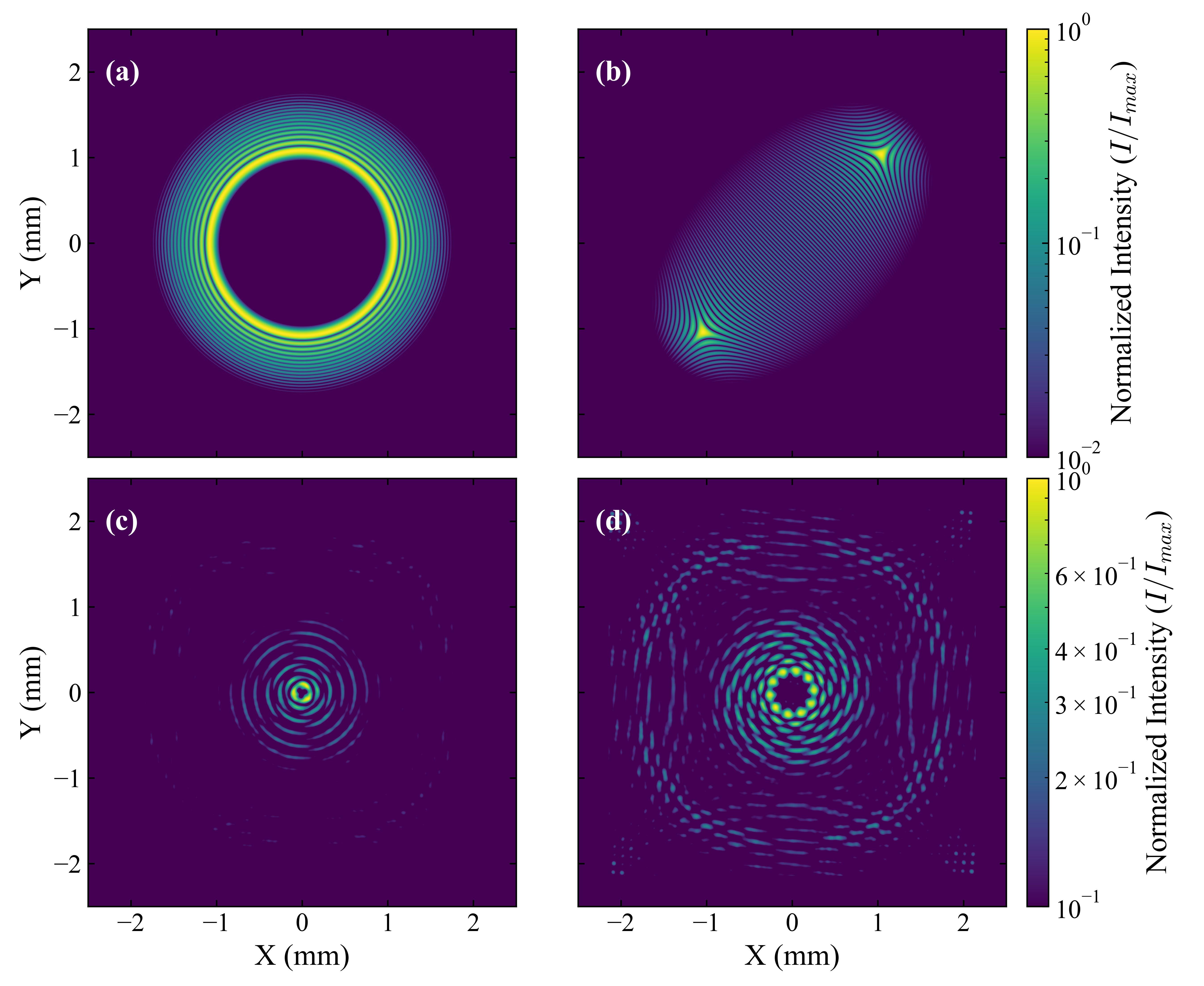}
        \caption{Simulation of structured light generation and transformation. 
        \textbf{(a)} Transformation of a Gaussian input into a Laguerre--Gaussian mode carrying the OAM $\ell = 64 \hbar$. 
        \textbf{(b)} Cylindrical-lens conversion into the corresponding Hermite--Gaussian mode of the order $N = 65$. 
        \textbf{(c)} Diffraction of a beam on a binary axicon with topological charge $m = 3$. 
        \textbf{(d)} Diffraction on a binary axicon with topological charge $m = 10$.}
        \label{fig:spp_axicon_sim}
    \end{figure*}

    This modeling procedure
    allowed us to directly compare the fork-grating–generated and SPP-generated beams, including verification of the expected OAM values. The 
    simulated output patterns are shown in Figures~\ref{fig:fork_sim} and \ref{fig:spp_axicon_sim}(a) and (d).

    Similarly to the Laguerre--Gaussian simulations, quasi--Bessel beams produced in the focal region of the fabricated axicons were numerically modeled following the methodology of 
    Ref.~\cite{Bazdyrev25}. The results are shown in Figures~\ref{fig:spp_axicon_sim}(c) and (d). For all topological charge values, the characteristic ring-shaped intensity distribution is reproduced, 
    consisting of $m$ localized bright lobes, the number of which is equal
    to the topological charge of each axicon. Small parasitic background features appear as weak spiral-like artifacts.

    We also note that the binary phase pattern can lead to an effective interference of several overlapping wave components carrying slightly different OAMs. 
    In this interpretation, the quasi--Bessel field can be viewed not as a single-mode vortex, but as a superposition of nearby OAM 
    states~\cite{Vasilyeu09, Kovalev15, Saadati-Sharafeh20}. Then, the average topological charge
    still corresponds to the design value $m$, while the OAM can exhibit a finite 
    dispersion~\cite{CARRUTHERS68}, defining an effective OAM bandwidth. This qualitative picture is consistent with the observed fragmented ring structure and is commonly discussed for binary 
    axicons.

    In contrast, an ideal (non-binary) axicon produces a continuous, smooth ring~\cite{Osintseva22}. In the present case, the discretization inherent to the binary phase 
    patterns results in a fragmented ring composed of distinct intensity maxima, consistent with the expected behavior of binary axicons.

\subsection{OAM spectrum decomposition}\label{sec:oam_spectrum}

    To quantify the OAM content of the generated fields, we project the numerically obtained complex transverse field onto azimuthal harmonics 
    $\exp(i\ell\varphi)$~\cite{Torner05,Vasnetsov05}. The projection coefficient is evaluated as

    \begin{equation}
        c_\ell \propto \int E(r,\varphi)\, \exp(-i\ell\varphi)\, W(r)\, r\, \mathrm{d}r\, \mathrm{d}\varphi,
    \end{equation}
    where $W(r)$ accounts for the finite aperture. The normalized spectrum is defined as

    \begin{equation}
        P_\ell = |c_\ell|^2 \qquad \sum_\ell P_\ell = 1.
    \end{equation}

\begin{figure}
        \centering
        \includegraphics[width=0.85\linewidth]{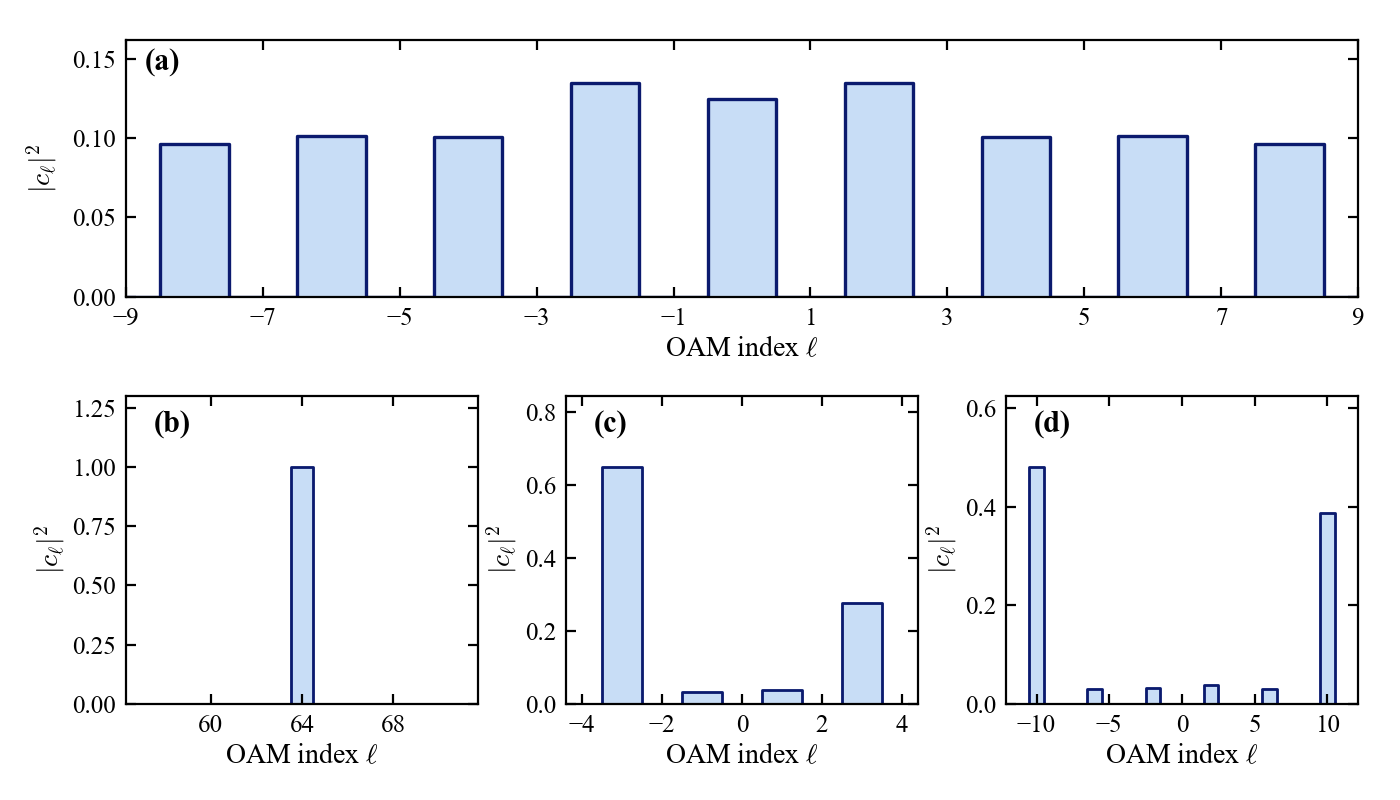}
        \caption{OAM
        spectrum of the generated beams obtained by numerical projection onto azimuthal harmonics. 
        \textbf{(a)} Fork grating: discrete, well-separated peaks at $\ell=\pm2 \hbar$, $\pm4 \hbar$, and $\pm8 \hbar$, corresponding to the individual diffraction orders of the grating and indicating 
        high OAM spectral purity. 
        \textbf{(b)} Spiral phase plate (SPP): a single narrow peak centered at the
        intended topological charge, demonstrating the generation of a nearly pure OAM eigenstate. 
        \textbf{(c)} Binary axicon with topological order $m=3$ and 
        \textbf{(d)} binary axicon with $m=10$: broadened and structured OAM spectra resulting from a superposition of multiple azimuthal harmonics with opposite topological charges, 
        characteristic of quasi--Bessel beams.}
        \label{fig:oam_spectrum}
    \end{figure}

    For Laguerre--Gaussian-like beams generated by fork gratings and
    SPPs, the field contains a single azimuthal phase factor $\exp(i\ell_0\varphi)$. Owing to 
    the orthogonality of azimuthal harmonics, only $\ell=\ell_0$ survives upon integration, yielding a narrow single-peaked OAM spectrum (Figure~\ref{fig:oam_spectrum}(a--b)). 
    In fork gratings, different diffraction orders carry
    different, well-defined topological charges, producing well-separated peaks at $\ell=\pm2 \hbar$, $\pm4 \hbar$, and $\pm8 \hbar$.

    In contrast, binary axicons generate quasi--Bessel beams with coupled radial and azimuthal
    features. The field can be represented as a superposition of opposite OAM 
    components $\pm\ell$ whose weights vary
    with radius~\cite{Sanchez-Lopez20,Fu22}. As a result, the azimuthal projection yields multiple non-vanishing components and a 
    broadened OAM spectrum with a finite intrinsic bandwidth (Figures~\ref{fig:oam_spectrum}(c) and (d)).

    To reduce numerical cancellation caused by rapid radial phase oscillations, the aperture is divided into $N$ concentric annuli, within which local projections $c_\ell^{(j)}$ 
    are evaluated and summed as

    \begin{equation}
         P_\ell=\sum_j |c_\ell^{(j)}|^2.
    \end{equation} 
    This procedure preserves the physical OAM content while improving the numerical stability.

\subsection{Control of mode purity in beams generated by a binary axicons}
    As shown in Figures~\ref{fig:oam_spectrum}(a) and (b), fork gratings and
    SPPs generate nearly pure OAM eigenstates. In contrast, binary spiral axicons produce 
    coherent superpositions of azimuthal harmonics, leading to intrinsically mixed OAM spectra. 

\begin{figure}[!h]
        \centering
        \includegraphics[width=\linewidth]{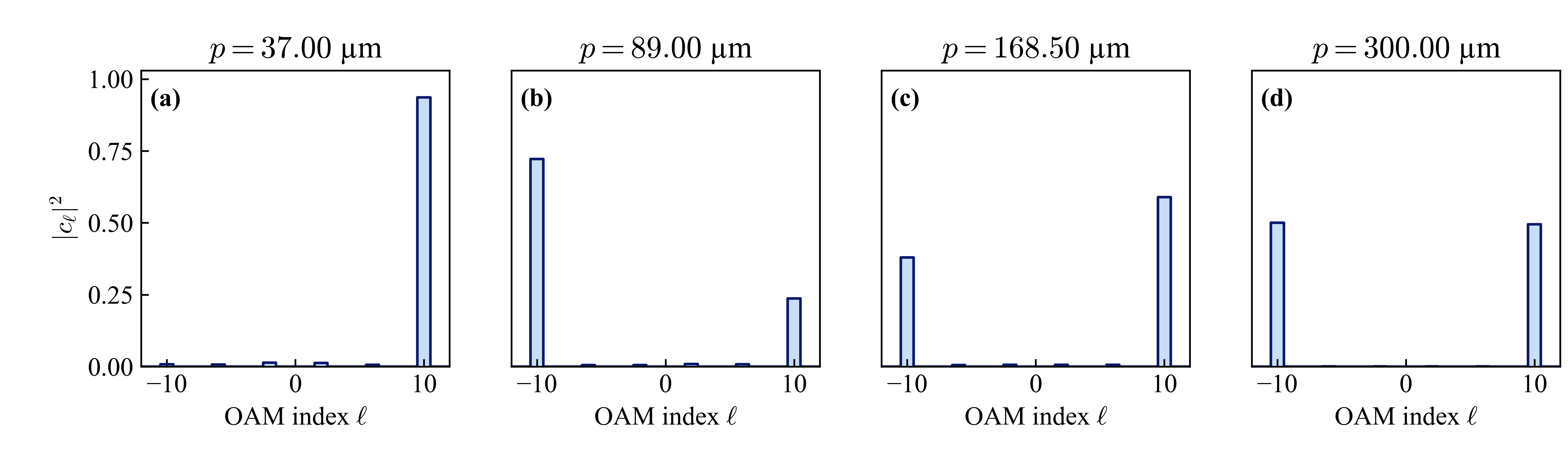}
        \caption{Projected OAM spectra $|c_\ell|^2$ of the field generated by
        binary spiral axicons with
        different transverse periods: 
        \textbf{(a)} $p = 37.00~\mu$m, 
        \textbf{(b)} $p = 89.00~\mu$m, 
        \textbf{(c)} $p = 168.50~\mu$m, and (d) $p = 300.00~\mu$m. A decrease of the period enhances the dominance of a single $\ell$ component, whereas larger periods lead to 
        pronounced mixing of opposite topological charges and the formation of symmetric $\pm \ell$ superpositions. All other parameters are kept fixed.}
        \label{fig:oam_control_p}
    \end{figure}

\begin{figure}[!h]
        \centering
        \includegraphics[width=0.7\linewidth]{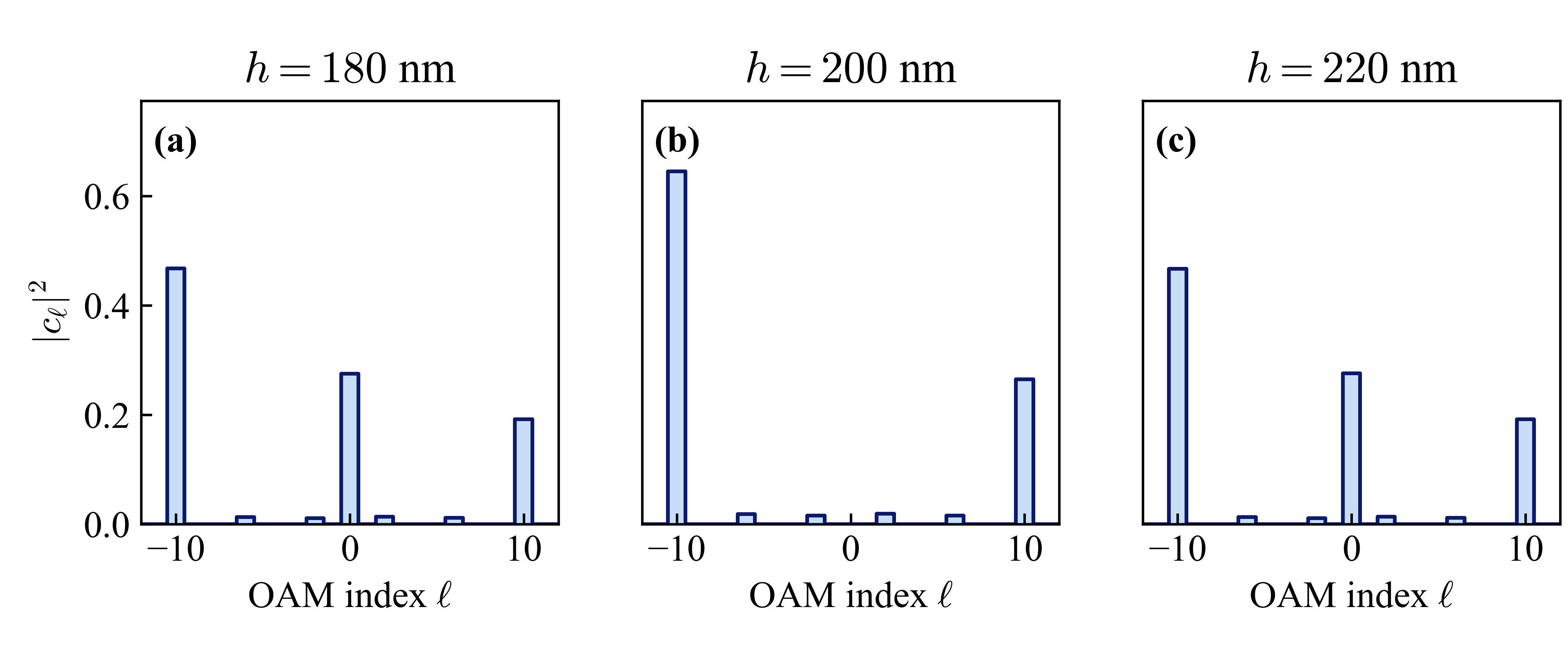}
        \caption{Projected OAM spectra $|c_\ell|^2$ of the field generated by the binary spiral axicon for different SU-8 thicknesses: 
        (a) $h = 180$ nm, 
        (b) $h = 200$ nm, and (c) $h = 220$ nm. 
        Variation of the phase modulation depth leads to the appearance of a non-negligible $\ell = 0$ component and a redistribution of the relative weights of the 
        $\pm\ell$ harmonics. All other parameters are kept fixed.}
        \label{fig:oam_control_h}
    \end{figure}

    The modal composition formed by an axicon can be tuned by varying its
    transverse period $p$ or the SU-8 layer thickness $h$ (Figures~\ref{fig:oam_control_p},~\ref{fig:oam_control_h}). Decreasing 
    $p$ leads to
    the dominance of a single $\ell$ component; however, the fabricated value $p=100~\mu$m 
    was chosen to obtain a focal spot of several hundred micrometers required for the photoemission experiment. A smaller period would increase the transverse spatial frequency 
    and substantially enlarge the focal spot size in our geometry. Consequently, in the deep-UV
    diffraction regime considered here, the axicon mostly generates 
    a superposition of pure OAM states with dominant $\pm\ell$ components or their symmetric superposition.

    Variation of the thickness $h$ modifies the effective complex transmission function of the axicon and
    induces a non-negligible $\ell = 0 \hbar$ 
    contribution in the OAM spectrum (Figure~\ref{fig:oam_control_h}). Even small $h$ changes of several tens of nanometers noticeably reduce the modal purity. %
    The reasons are, the phase modulation deviates from its optimal value, and an
    amplitude modulation is introduced by the finite conductivity of the metal. In the experiment, 
    such a contribution would manifest
    as a Gaussian-like on-axis intensity core superimposed on the annular Bessel structure.

\newpage

\section{Experimental results and beam imaging}\label{sec:experiment}

\subsection{Experimental setup}\label{sec:setup}

    \begin{figure}[!ht]
        \centering
        \includegraphics[width=\linewidth]{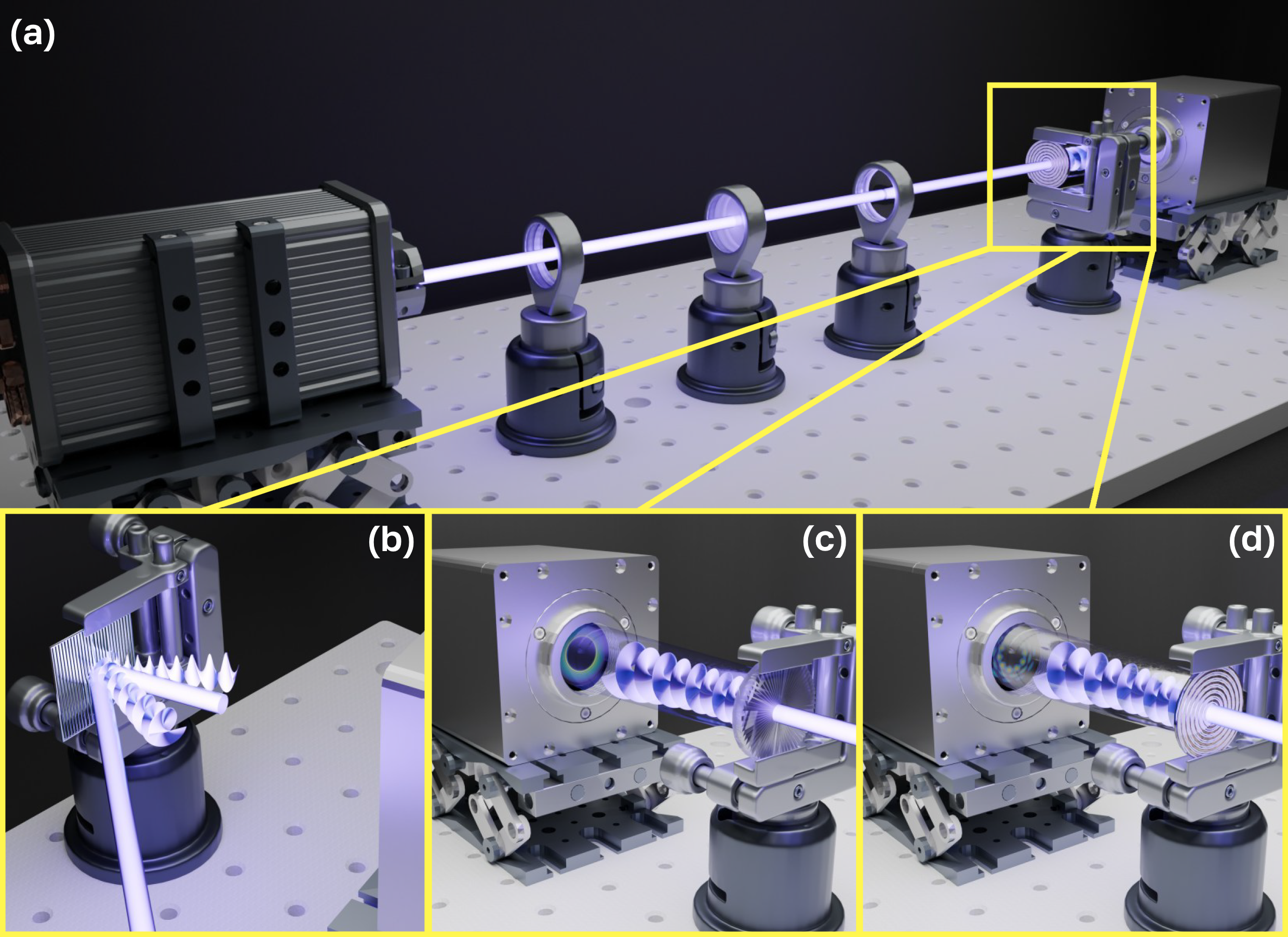}
        \caption{Schematic layout of the experimental setup and representative beam-formation configurations.
        \textbf{(a)} General view of the optical experiment, including the drive laser beamline, beam-conditioning optics, and the detection system.
        \textbf{(b)} Diffraction of the laser beam by a reflective fork grating, yielding Laguerre--Gaussian modes in several diffraction orders with topological charges proportional 
        to the grating topological charge $m = 2$.
        \textbf{(c)} Generation of twisted light carrying OAM
        $\ell = 64\hbar$ using an SPP.
        \textbf{(d)} Formation of a coherent superposition of quasi--Bessel twisted light modes produced by a binary axicon.}
        \label{fig:general_scheme}
    \end{figure}

    Figure~\ref{fig:general_scheme} shows the optical setup used to characterize the fabricated DOEs. The
    UV beam with a
    wavelength of $\lambda = 260$--$266$~nm was 
    extracted from the photoinjector transport line using a secondary periscope and directed through a spatial filtering stage consisting of a focusing lens and a $200\ \mu$m pinhole.
    In the current configuration, the initial driving laser beam is linearly polarized, and the subsequent transmissive DOEs largely preserve this macroscopic polarization state. However, since 
    linear polarization is fundamentally a coherent superposition of left- and right-handed circular polarizations, the resulting vortex beam contains a mixture of spin angular momentum states. To generate the 
    purest possible twisted mode with a strictly defined total angular momentum, a quarter-wave ($\lambda/4$) plate can be easily integrated into the optical layout before the DOEs. Converting the 
    incident light to a single circular polarization state would eliminate spin-mode degeneracy, providing an optimal pure optical state for fundamental photoemission studies.
    For measurements involving the fork grating, an additional focusing lens was introduced to provide tighter illumination of the DOE.
    
    After spatial filtering, the beam illuminated the DOE under test (fork grating, SPP, or axicon). For the OAM diagnostics in the fork-grating and SPP experiments, a 
    cylindrical lens with a focal length of $100$ mm was used to perform Laguerre--Gaussian to Hermite--Gaussian (LG-to-HG) mode conversion~\cite{Beijersbergen92, Kotlyar17}. 
    The resulting intensity distributions were recorded with a UV-sensitive SDU-285R CCD camera.

    This optical layout can also form
    a Gaussian beam with a transverse diameter of approximately $800\ \mu\mathrm{m}$ and low angular divergence, suitable for 
    subsequent illumination of photocathodes or additional diagnostics.

\subsection{Experiments with fork grating}
\label{sec:fork_grat_exp}

\begin{figure*}[!ht]
        \centering
        \begin{minipage}{0.48\linewidth}
            \centering
            \begin{overpic}[width=\linewidth]{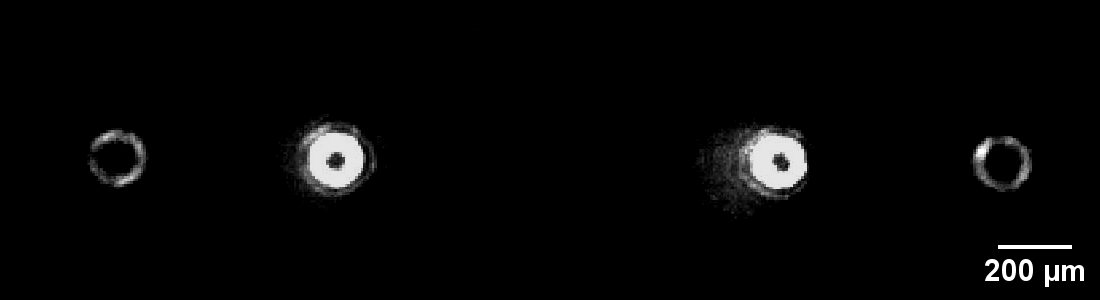}
                \put(2,22){\colorbox{black}{\textcolor{white}{\bfseries\small (a)}}}
            \end{overpic}\vspace{2mm}
            \begin{overpic}[width=\linewidth]{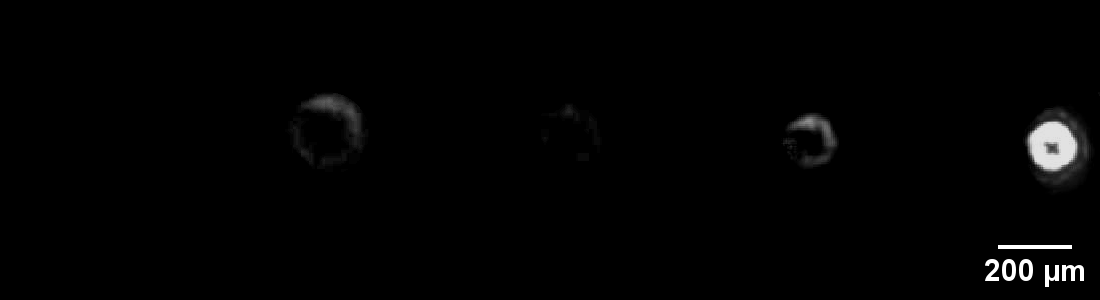}
                \put(2,22){\colorbox{black}{\textcolor{white}{\bfseries\small (c)}}}
            \end{overpic}
          \end{minipage}
          \hspace{2mm}
          \begin{minipage}{0.48\linewidth}
            \centering
            \begin{overpic}[width=\linewidth]{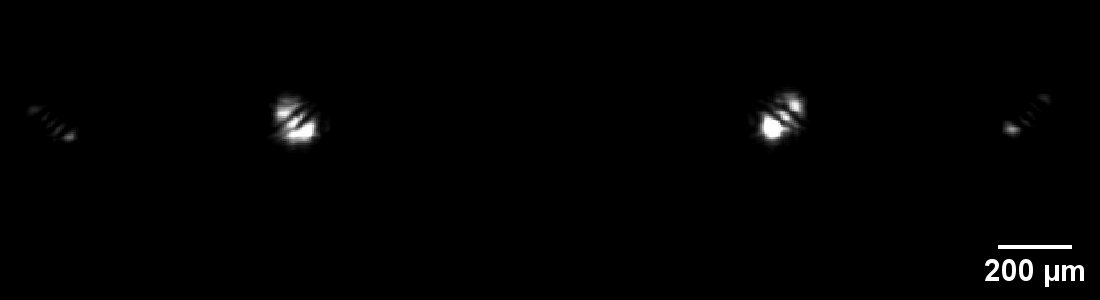}
                \put(2,22){\colorbox{black}{\textcolor{white}{\bfseries\small (b)}}}
            \end{overpic}\vspace{2mm}
            \begin{overpic}[width=\linewidth]{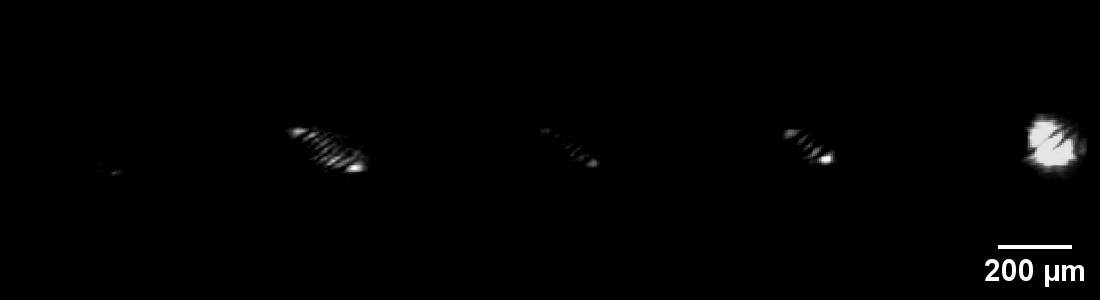}
                \put(2,22){\colorbox{black}{\textcolor{white}{\bfseries\small (d)}}}
            \end{overpic}
        \end{minipage}
        \caption{
        Twisted light generated by the reflective fork grating.
        (a) Laguerre--Gaussian modes with the OAM $\ell=\pm2$ and $\ell=\pm4$ observed in the $\pm1$ and $\pm2$ diffraction orders.
        (b) Corresponding Hermite--Gaussian modes obtained after cylindrical-lens mode conversion.
        (c) Higher-order Laguerre--Gaussian modes with $\ell=-2,\,-4,\,-6,\,-8$ observed in the $n=-1,-2,-3,-4$ diffraction orders.
        (d) Corresponding Hermite--Gaussian modes obtained after cylindrical-lens mode conversion.
        }
        \label{fig:fork_grat_exp}
    \end{figure*}

    The reflective fork grating was first tested using the UV beam prepared by the spatial
    filtering stage. 
    The generated twisted light beams are shown in Figure~\ref{fig:fork_grat_exp}. In these measurements, the central (zeroth-order) beam was blocked to avoid saturation of the CCD 
    sensor. As expected for a grating with $m = 2$, the $\pm 1$ and $\pm 2$ diffraction orders produced Laguerre--Gaussian modes with
    OAM values $\ell = \pm 2 \hbar$ and 
    $\ell = \pm 4 \hbar$, respectively. Power measurements using an Ophir 3A-FS detector confirmed that
    most of the transmitted energy was contained in the zeroth and first 
    diffraction orders, while higher orders had low intensities.

    The corresponding Hermite--Gaussian modes obtained after cylindrical-lens conversion are presented in Figure~\ref{fig:fork_grat_exp}(b). The observed modal structures agree 
    with the expected relation~\cite{Beijersbergen92, Kotlyar17}

    \begin{equation}
        N = |\ell| + 1,
    \end{equation}
    demonstrating that the grating reliably generates the
    desired OAM states. 
    Higher diffraction orders ($n = -3$ and $n = -4$), shown in Figures~\ref{fig:fork_grat_exp}(c) and (d), also exhibit well-defined modal patterns, confirming correct operation 
    of the fabricated structure across multiple orders.

    To improve the performance of a diffractive element, the spatial resolution of the grating lines could be increased. 
    Indeed, a finer grating would generate higher-quality vortex modes, but it would also require tighter focusing of the incident beam on
    the grating surface. 
    Overall, the measured spatial profiles in Figures~\ref{fig:fork_grat_exp}(a) and (b) show good qualitative agreement with the simulations presented 
    in Figure~\ref{fig:fork_sim}, validating the fabricated fork grating.

\subsection{Experiments with SPP}
\label{sec:spp_exp}

    Next, we tested
    the fabricated SPP, and the resulting beam profiles are shown in Figures~\ref{fig:twisted_light_exp}(a) and (b). The output twisted 
    light (see Figure~\ref{fig:twisted_light_exp}(a)) exhibits a well-defined annular structure with a pronounced central singularity. A portion of the beam energy is redistributed into 
    several weak concentric rings surrounding the primary vortex mode, consistent with the expected behavior of high-charge SPP-generated beams.

    Power was measured with the Ophir 3A-FS detector, showing the beam intensity reduced by approximately $20\%$ after the SPP. 
    Although this differs from the theoretical SPP conversion 
    efficiency of $\sim 90\%$, the obtained value is still reasonable given the surface-profile discretization, potential scattering losses, and alignment constraints in the UV range. 
    The measured twisted light structure and transverse size closely match the simulated pattern shown in Figures~\ref{fig:spp_axicon_sim}(a) and (b).

    The Hermite--Gaussian mode (see Figure~\ref{fig:twisted_light_exp}(b)) exhibits a slightly larger tilt compared with the simulation.
    The reason is, the beam is intentionally slightly misaligned relative to the focal plane of the cylindrical lensd 
    to improve the separation and visibility of the experimentally observed 
    mode stripes.

\subsection{Experiments with binary axicons}
\label{sec:axicon_exp}
    
\begin{figure*}[!ht]
        \centering
        \begin{minipage}[t]{0.7\linewidth}
            \centering
            \begin{overpic}[width=0.48\linewidth]{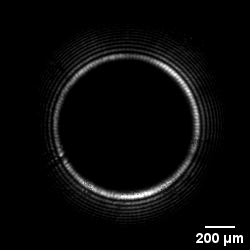}
                \put(2,5){\colorbox{black}{\textcolor{white}{\bfseries\small (a)}}}
            \end{overpic}\hspace{2mm}
            \begin{overpic}[width=0.48\linewidth]{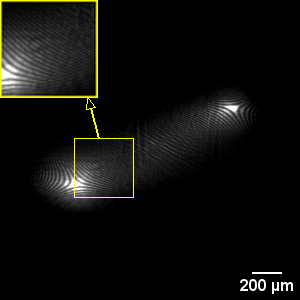}
                \put(2,5){\colorbox{black}{\textcolor{white}{\bfseries\small (b)}}}
            \end{overpic}
        \end{minipage}
        \vspace{2mm}
        \begin{minipage}[t]{0.7\linewidth}
            \centering
            \begin{overpic}[width=0.48\linewidth]{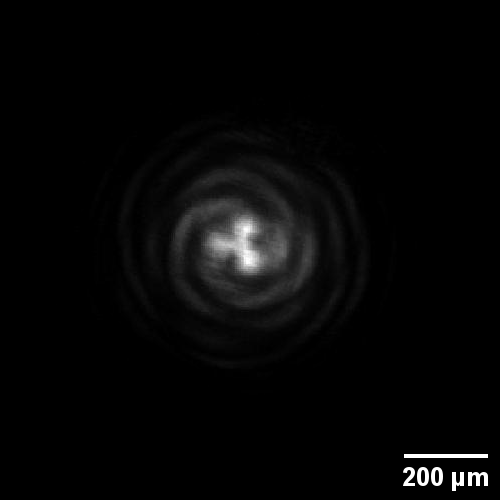}
                \put(2,5){\colorbox{black}{\textcolor{white}{\bfseries\small (c)}}}
            \end{overpic}\hspace{2mm}
            \begin{overpic}[width=0.48\linewidth]{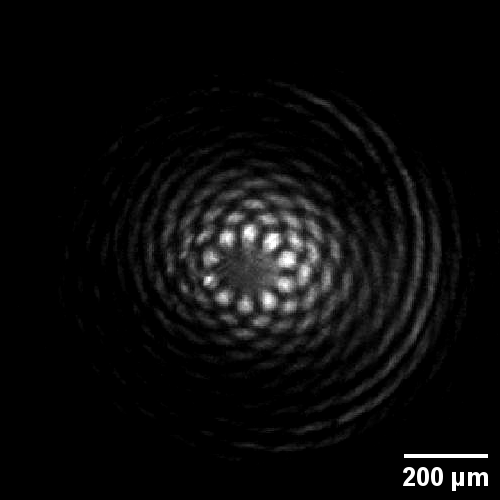}
                \put(2,5){\colorbox{black}{\textcolor{white}{\bfseries\small (d)}}}
            \end{overpic}
        \end{minipage}
        \caption{
        Twisted light generated by structured optical elements.
        (a) Laguerre--Gaussian vortex carrying the OAM $\ell = 64$ generated by the SPP.
        (b) Corresponding Hermite--Gaussian mode of the order $N = 65$ obtained after
        cylindrical-lens mode conversion.
        (c) Quasi--Bessel beam produced by the binary axicon with the topological charge
        $m = 3$, exhibiting three intensity lobes.
        (d) Quasi--Bessel beam produced by the binary axicon with the topological charge
        $m = 10$, exhibiting ten resolved lobes.
        }
        \label{fig:twisted_light_exp}
    \end{figure*}

    Finally, we experimentally investigated the fabricated binary axicons. The resulting quasi--Bessel beams are shown in Figures~\ref{fig:twisted_light_exp}(c) and (d). 
    For the axicon with the topological charge $m = 3$ (see Figure~\ref{fig:twisted_light_exp}(c)), three intensity lobes are clearly visible,
    as expected. At short propagation distances, these lobes partially overlap and form a characteristic Y-shaped pattern, 
    typical for binary axicons due to 
    the limited spatial separation between the initial diffraction maxima.

    In contrast, the axicon with $m = 10$ produces a well-resolved multi-lobe ring (see Figure~\ref{fig:twisted_light_exp}(d)), in good agreement with the simulated patterns 
    in Figures~\ref{fig:spp_axicon_sim}(c) and (d). Both axicons exhibit weak spiral-like background features associated with the binary phase discretization.

    Measured power losses did not exceed $30\%$, indicating high transmission and good mode quality. As the propagation distance increased, the ring radius expanded while the 
    internal lobe structure remained nearly unchanged, consistent with the expected behavior of quasi--Bessel beams with low angular divergence.
    
    Finally, we note that Bessel and quasi--Bessel beams cannot be converted into Hermite--Gaussian modes using a cylindrical-lens scheme, since their propagation-invariant 
    structure does not undergo the conventional LG-to-HG transformation. For such beams, the OAM
    projection simply coincides with the topological charge 
    $m$. As shown in Ref.~\cite{Knyazev23}, $m$ can be obtained simply by counting the number of intensity maxima in the ring, providing a straightforward and reliable method 
    for identifying the OAM content of quasi--Bessel beams. In practice, quasi--Bessel beams can exhibit a finite OAM bandwidth due to small contributions from the nearby OAM 
    states, but the mean topological charge still equals to $m$, as discussed in Section~\ref{sec:num_sim}. 

\section{Discussion and conclusion}\label{sec:conclusion}

    \begin{figure}[!ht]
        \centering
        \includegraphics[width=\linewidth]{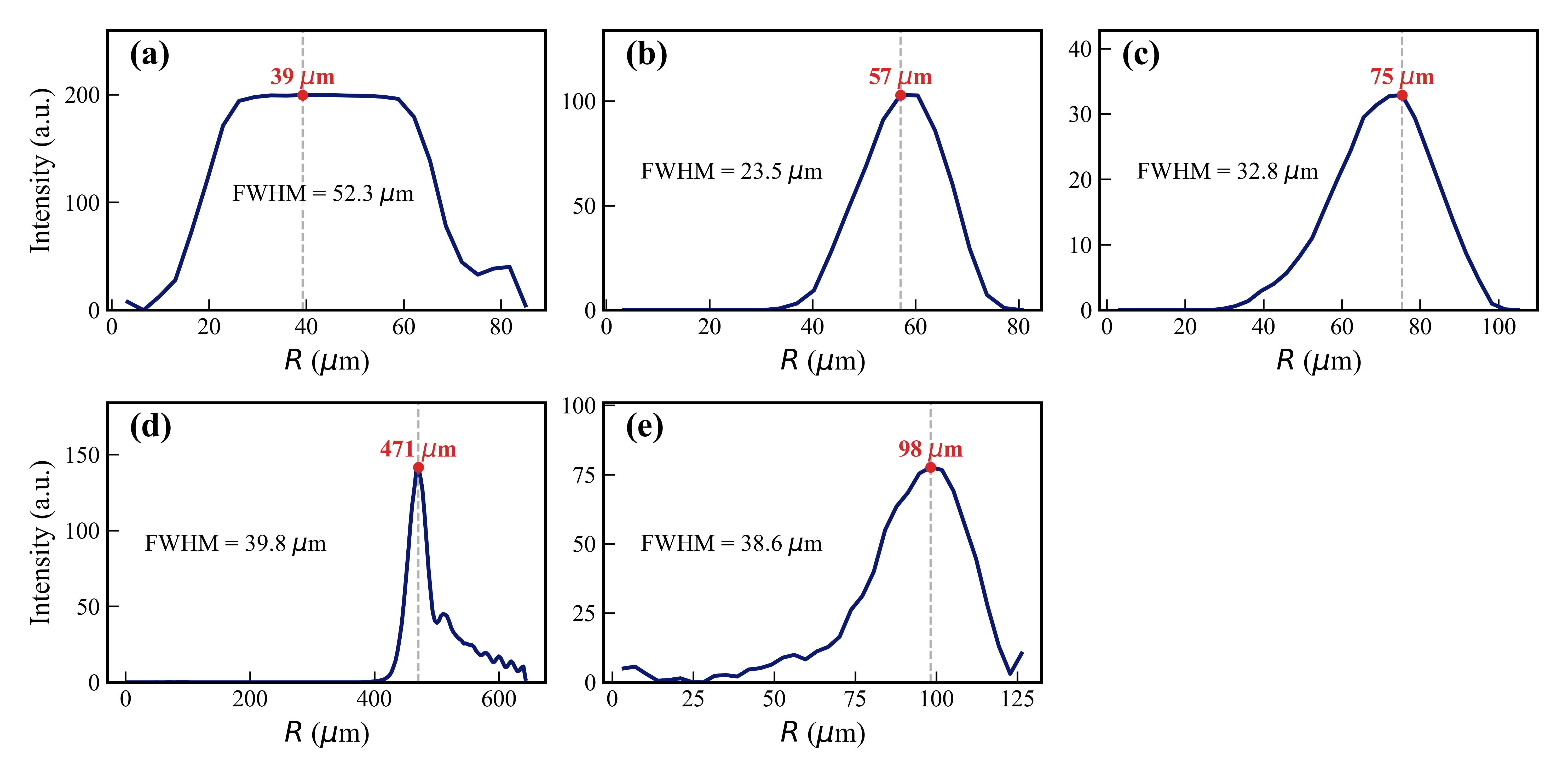}
        \caption{Radial profiles of the generated twisted light beams:
        \textbf{(a)} -- \textbf{(c)} -- fork grating producing the OAM $\ell=2$, $\ell=4$ and $\ell=8$ modes;  
        \textbf{(d)} -- SPP generating a high-order LG beam with $\ell=64$;  
        \textbf{(e)} -- binary axicon producing a quasi--Bessel beam with the topological charge $m=10$.}
        \label{fig:radial_profiles}
    \end{figure}

    We have experimentally demonstrated the generation and 
    propagation of optical beams carrying orbital angular momentum (OAM) in the deep-ultraviolet (UV) range using fabricated diffractive 
    optical elements, including fork gratings, spiral phase plates (SPPs), and binary axicons. A consistent angular-spectrum–based modeling approach was used to compare the resulting 
    transverse intensity distributions and modal properties.

    Fork gratings generate Laguerre--Gaussian modes with systematically increasing ring radii and well-defined annular profiles,\ 
    with the experimental radial 
    distributions shown in Figures~\ref{fig:radial_profiles}(a)--(c) and the simulations in Figure~\ref{fig:fork_sim}.
    SPPs provide the highest modal purity and conversion efficiency, stably generating high-order OAM states up to $\ell = 64 \hbar$, as illustrated in 
    Figures~\ref{fig:radial_profiles}(d) and \ref{fig:spp_axicon_sim}(a), (b). 
    
    It should be noted that $\ell = 64 \hbar$ is not a theoretical upper limit for our methods, but rather a practical optimum dictated by experimental and fabrication 
    constraints. For the reflective fork grating, where the initial Gaussian beam is focused onto the element, the maximum achievable OAM is restricted by the beam spot size, which 
    limits the number of illuminated grating periods. For the SPP, the limitation arises directly from the lithographic resolution and the maximum feasible etching depth. For instance, 
    based on the phase matching condition, to generate
    an OAM of $\ell = 100 \hbar$ at $\lambda = 266$~nm, an SPP of
    fused silica ($n \approx 1.49$) would require 100 azimuthal sectors and a total 
    relief height of $h_s = 100 \lambda / (n - 1) \approx 54.3\ \mu$m. Fabricating such a deep profile while maintaining sharp vertical walls between the sectors is technologically 
    challenging. Furthermore, the lateral width of each sector at a radial distance $r$ is $2\pi r / 100$; near the optical axis, this dimension falls below typical lithographic 
    resolution limits (e.g., $\sim 1\ \mu$m), leading to a defective vortex core. We therefore chose $\ell = 64 \hbar$ as a representative example to demonstrate successful high-order OAM 
    generation in the deep UV while strictly
    preserving high beam fidelity.
    
    Binary axicons produce low-divergence quasi–Bessel beams with segmented ring-shaped intensity profiles and 
    a finite OAM bandwidth corresponding to a superposition of multiple OAM states,
    with the experimental and numerical results in Figures~\ref{fig:radial_profiles}(e) 
    and \ref{fig:spp_axicon_sim}(c), (d). A comparative overview of the three diffractive elements is summarized in Table~\ref{tab:comparison}.

    The demonstrated approach enables the generation of high-order OAM beams up to $\ell = 64 \hbar$ with well-defined transverse structure, suitable for OAM transfer to relativistic electrons~\cite{Pavlov24} 
    and diagnostics in RF photoinjectors. By illuminating the photocathode with these macroscopic twisted beams, our ultimate scientific goal is the generation of relativistic vortex electron 
    beams carrying quantized OAMs. Such electron beams are of immense interest for fundamental physics and advanced beam diagnostics. Specifically, highly relativistic vortex electrons hold great potential 
    for high-energy physics, including novel approaches to analyzing the proton spin structure~\cite{Ivanov26}. Furthermore, precise control over the transverse phase of the electron wave function is highly 
    relevant for the development of a quantum electron microscope~\cite{Kruit16, Inbar23}, enabling non-destructive imaging and enhanced phase contrast. 
    
    Since the divergence of Laguerre--Gaussian modes scales as $\sqrt{\ell}$~\cite{Karlovets21}, beamlines for high-OAM injection must account for the corresponding increase of the Twiss $\beta$-function and 
    transverse emittance~\cite{Wiedemann15,Chao20,Lee21}. For metallic photocathodes, SPPs and axicons are particularly attractive due to their higher efficiency, modal purity, and power handling. The high 
    modal purity of the SPP beam also facilitates OAM diagnostics based on LG--HG conversion using astigmatic optics~\cite{Schattschneider12}, while diffraction on a triangular aperture provides a viable 
    alternative for quasi--Bessel beams generated by axicons~\cite{Maksimov25}. Overall, the presented results establish a direct connection between diffractive phase design, numerical propagation, and 
    structured light generation in the deep-UV regime, relevant for photocathode-based RF photoinjectors and accelerator-driven applications.
    
    Furthermore, while our current study focuses on integer topological charges tailored for specific photoinjector dynamics, this approach can be extended to fractional 
    topological charges in future research. Unlike integer modes, optical beams with fractional OAM exhibit an asymmetric transverse intensity profile (typically a C-shaped distribution) 
    and a complex line of phase singularity. If applied to a photocathode, such beams would generate rotationally asymmetric electron bunches with a highly non-trivial initial transverse 
    phase space. This structured emission provides an additional degree of freedom for advanced beam manipulation techniques, such as precise transverse emittance partitioning and phase-space 
    shaping prior to acceleration. Moreover, recent advances in the robust measurement of fractional topological charges in optics~\cite{Shikder24} provide a crucial diagnostic foundation. 
    These optical methodologies could inspire analogous diagnostic techniques for characterizing asymmetric electron distributions, offering new tools for probing complex beam evolution in high-brightness 
    accelerator environments.

    \begin{table}[!ht]
      \centering
      \begin{tabular}{l c c c}
            \hline
            Parameter & Fork Grating & SPP & Axicon \\
            \hline
            Fabrication complexity & Low & Very High & High \\
            DUV compatibility & Good & Very High & Very High \\
            Efficiency & 60\% & 80\% & 70\% \\
            Cost & Low & Very High & Mid \\
            Mode Quality & High & Very High & High \\
            Suggested Photocathode Type & Semiconductor & Any Type & Any Type \\
            \hline
      \end{tabular}
      \caption{Comparison of parameters for the fork grating, the spiral phase plate, and the binary axicon.}
      \label{tab:comparison}
    \end{table}

\medskip
\textbf{Acknowledgements} \par 
    We deeply acknowledge A.A.~Bogdanov for the help that made the fabrication of the spiral phase plate possible. We sincerely thank K.V.~Cherepanov for invaluable assistance 
    and advice. We are also grateful to S.S.~Baturin for helpful discussions and constructive feedback. We additionally thank N.E.~Sheremet for assistance with developing the 
    numerical code used to generate the fork-grating phase masks. The axicons were simulated and fabricated at the shared research facility of Siberian Center for Synchrotron 
    and Terahertz Radiation at the basis of the Novosibirsk Free Electron Laser, Budker Institute of Nuclear Physics SB RAS. The authors sincerely thank Maria Zhuravleva for 
    her valuable assistance in preparing the graphical materials for this work.
    
    This study was supported by the Russian Science Foundation (Project No. 23-62-10026)~\cite{RSF}.

\medskip
\textbf{Conflicts of Interest} \par
The authors declare no conflicts of interest.

\medskip
\textbf{Data Availability Statement} \par
The data that support the findings of this study are available from the corresponding author upon reasonable request.

\bibliographystyle{elsarticle-num} 
\bibliography{refs}

\end{document}